\documentclass[
    ,final            
  ]
  {aipproc}

\layoutstyle{6x9}
\usepackage{amsmath,aas_macros}

\begin{document}

\title{The Detectability of AGN Cavities in Cooling-Flow Clusters}
 

\author{L. B\^{\i}rzan}{
  address={Department of Astronomy and Astrophysics, Pennsylvania State University, 525 Davey Lab., University Park, PA 16802, USA\footnote{Current address: Leiden Observatory, Leiden University, Oort Gebouw, P.O. Box 9513, 2300 RA Leiden, The Netherlands}} 
}

\author{D. A. Rafferty}{
  address={Department of Astronomy and Astrophysics, Pennsylvania State University, 525 Davey Lab., University Park, PA 16802, USA\footnote{Current address: Leiden Observatory, Leiden University, Oort Gebouw, P.O. Box 9513, 2300 RA Leiden, The Netherlands}} 
}

\author{B. R. McNamara}{
  address={Department of Physics and Astronomy, University of Waterloo, Waterloo, ON N2L 2G1, Canada}
  ,altaddress={Perimeter Institute for Theoretical Physics, Waterloo, ON N2L 2Y5, Canada} 
  ,altaddress={Harvard-Smithsonian Center for Astrophysics, 60 Garden St., Cambridge, MA 02138, USA} 
}

\author{P. E. J. Nulsen}{
  address={Harvard-Smithsonian Center for Astrophysics, 60 Garden St., Cambridge, MA 02138, USA}
 }
 
 \author{M.~W.~Wise}{
  address={Astronomical Institute Anton Pannekoek, University of Amsterdam, Kruislaan 403, 1098 SJ Amsterdam, The Netherlands}
 }

\begin{abstract}

\emph{Chandra} X-ray Observatory has revealed X-ray cavities in many nearby cooling flow clusters. The cavities trace feedback from the central active galactic nulceus (AGN) on the intracluster medium (ICM), an important ingredient in stabilizing cooling flows and in the process of galaxy formation and evolution. But, the prevalence and duty cycle of such AGN outbursts is not well understood. To this end, we  study how the cooling is balanced by the cavity heating for a complete sample of clusters (the Brightest 55 clusters of galaxies, hereafter B55). In the B55, we found 33 cooling flow clusters (with $t_{\rm{cool}}<3 \times  10^{9}$ yr in the core), 20 of which have detected X-ray bubbles in their ICM.  Among the remaining 13, all except Ophiuchus could have significant cavity power yet remain undetected in existing images. This implies that the duty cycle of AGN outbursts with significant heating potential in cooling flow clusters is at least 60\% and could approach 100\%, but deeper data is required to constrain this further.

\end{abstract}

\maketitle


\section{Introduction}

Images from the \emph{Chandra} X-ray Observatory  show that many clusters have X-ray cavities in their atmospheres (only a small number of cavities were known from previous X-ray observatories, e.g., in M87 and Perseus). The X-ray cavities are interpreted as bubbles created by the central AGN, which rise buoyantly through the ICM from the center of the cluster. They are filled with relativistic particles and magnetic field which emit synchrotron radiation visible at radio wavelengths. The energy that the cavities release in the ICM may be important for balancing cooling \citep{mcna00,fabi00,blan01} and in the process of galaxy formation and evolution \citep{crot06}, but is not clear whether the cavities are always present when heating is required. \citet{birz04,dunn04,dunn05,raff06} analyzed samples of cooling flow clusters with visible cavities in their environments. \citet{raff06} found that  50-80 \% of the systems with cavities can balance cooling, considering only the enthalpy ($4pV$for $\gamma=4/3$). 

Currently, the cavities are primarily detected by eye. Some of them have rims (e.g., A2052, \citet{blan01}) or occupy a large fraction of the emitting volume of the ICM (e.g., MS 0735+7421, \citet{mcna05}), which make them "clear" cases. However, the detectability of a bubble depends on its location, orientation \citep{enss02,brugg09}, its angular size and the depth of the observation.  As a result, we probable are missing some bubbles in the existing images. 

Our goal is to understand the biases and selection effects in the detectability of current X-ray cavity samples. Since, we do not find bubbles in all cooling flow clusters (however, some of the images are shallow), we want to investigate if these systems may have enough cavity power to balance cooling and yet remain undetected in current images.

\section{A complete sample}

Now after almost 10 years of \emph{Chandra} observations there are enough data in the archive to study the heating versus the cooling \citep{birz04,raff06} for a complete sample of clusters. In this work we present an analysis  the B55 sample \citep{edge90} since this is  one of the most observed complete sample of clusters of galaxies. B55 is a 2-10 keV flux limited sample. This sample was studied by \citet{dunn06} in order to determine the fraction of cooling flow clusters with bubbles. In order to separate the cooling flow clusters from the non-cooling flow clusters \citet{dunn06} used ROSAT data. They define a cluster to be a cooling flow cluster if $t_{\rm{cool}}<3 \times  10^{9}$ yr and if it had a large temperature drop, $T_{\rm{outer}}/T_{\rm{center}}>2$. They found that 14 out of 20 cooling flow clusters have clear bubbles. 

In order to separate the cooling flow clusters from non-cooling flow clusters we used \emph{Chandra} archive data. We used the same central cooling time cut-off as \citet{dunn06} of 3 $\times  10^{9}$ yr . This cooling time cut-off is in agreement with findings of \citet{raff08}, who found that in systems with visible cavities, the central cooling time is less than $3\times 10^{9}$  yr. Under this criterium, we found that 33 out of an initial 54 clusters are cooling flow clusters (\emph{Chandra} data for A3532 are still proprietary) . 20 systems out of the 33 cooling flow clusters have cavities: A496, A2204\citep{sand09}, A2063 \citep{kano06}, and another 17 systems that were published in \citet{raff06} sample of clusters with cavities. For the remaining 13 clusters we perform simulations in order to find the location and the size of bubbles that can be present in the image and remain undetected.

\section{Simulations}

We simulate a 3 dimensional cluster using a double-$\beta$ profile for the emissivity (derived from the existing archive observations). Then, the cavities are subtracted and by integrating the emissivities along the line of sight, the 2 dimensional surface brightness profile is obtained. This profile is used as input for the MARX simulator\footnote{see \url{http://www.space.mit.edu/CXC/MARX/}} in order to obtain the simulated \emph{Chandra} image.

We use the adiabatic expansion assumption to put limits on the bubble size and location and the buoyancy assumption to calculate the ages. In order to calculate the radius at which the bubbles are injected we assumed: $4pV \sim P_{\rm{cav}}t \sim L_{\rm{X}}t$, where $p$ is the central pressure,  $L_{\rm{X}}$ is the bolometric X-ray luminosity and $t$ is the time between the outbursts. We adopted a time between outbursts of $10^{8}$ yr, motivated by the observations of clusters with multiple generation of bubbles, such as Perseus \citep{fabi00}.  We ran multiple simulations for a cluster, with the locations of the bubbles and the angle in the plane of sky ($\phi$) randomized, and with the angle from the line of sight ($\theta$) of 90$^\circ$. The resulting images are inspected by eye, and for the location at which the bubble becomes invisible ($R$) we check the age. If the buoyancy age ($t_{\rm{buoy}}$) is less then  $10^{8}$ yr, then such a bubble is physically possible under our assumptions and could then be present in the real cluster and yet remain undetected. If $t_{\rm{buoy}} >10^{8}$ yr, then there should be a new set of bubbles at smaller $R$ which would be detected, so this case is unphysical. Therefore, we repeated the simulation for smaller $P_{\rm{cav}}$ until $t_{\rm{buoy}} <10^{8}$ yr. This was the case for A2244, A2142 and Ophiuchus. Figure 1 left shows the simulated \emph{Chandra} image of A2244 for $\theta=90 ^{\circ}$, with much bigger bubbles than required to balance cooling (to make it easy to see the bubbles).

\section{Heating versus cooling for the B55 sample}

Figure 1 right shows the cavity power (the heating) versus the bolometric X-ray luminosity within the cooling radius (at which $t_{\rm{cool}}<7.7 \times  10^{9}$ yr ). The upper limits in the plot are the cooling flow systems without detected bubbles (13 of them) for which we performed simulations.  For the systems with detected bubbles we used the published values (except A496, for which we used our own calculations). From Figure 1 right we can conclude that most systems could have bubbles with enough power to balance cooling and remain undetected in existing images. One system (Ophiuchus) can not have such bubbles, unless they are at $\theta<30 ^{\circ}$. The fraction of systems with detected bubbles ($20/33\approx 0.6$) implies a duty cycle of at least 60\%, in rough agreement with the lower limit found by \citet{dunn06} in a similar study. However, with existing data, one can not exclude the scenario that all cooling flow clusters have bubbles with enough power to balance cooling, implying a duty cycle for such activity of up to 100\%. Much further work is needed to improve constraints on the duty cycle of cavity production, such as: adding rims to cavities, investigating different schemes for bubble evolution, developing an automated tool for detection and measurement, and quantifying the detection threshold. 

\begin{figure}
$\begin{array}{@{\hspace{0cm}}cc}
 \includegraphics[height=.3\textheight]{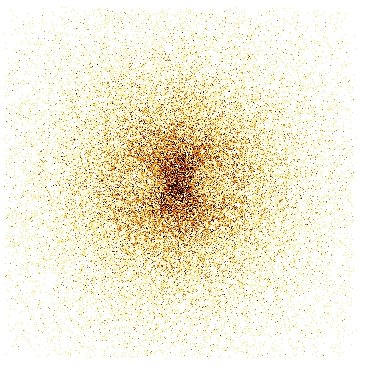}
  \includegraphics[height=.3\textheight]{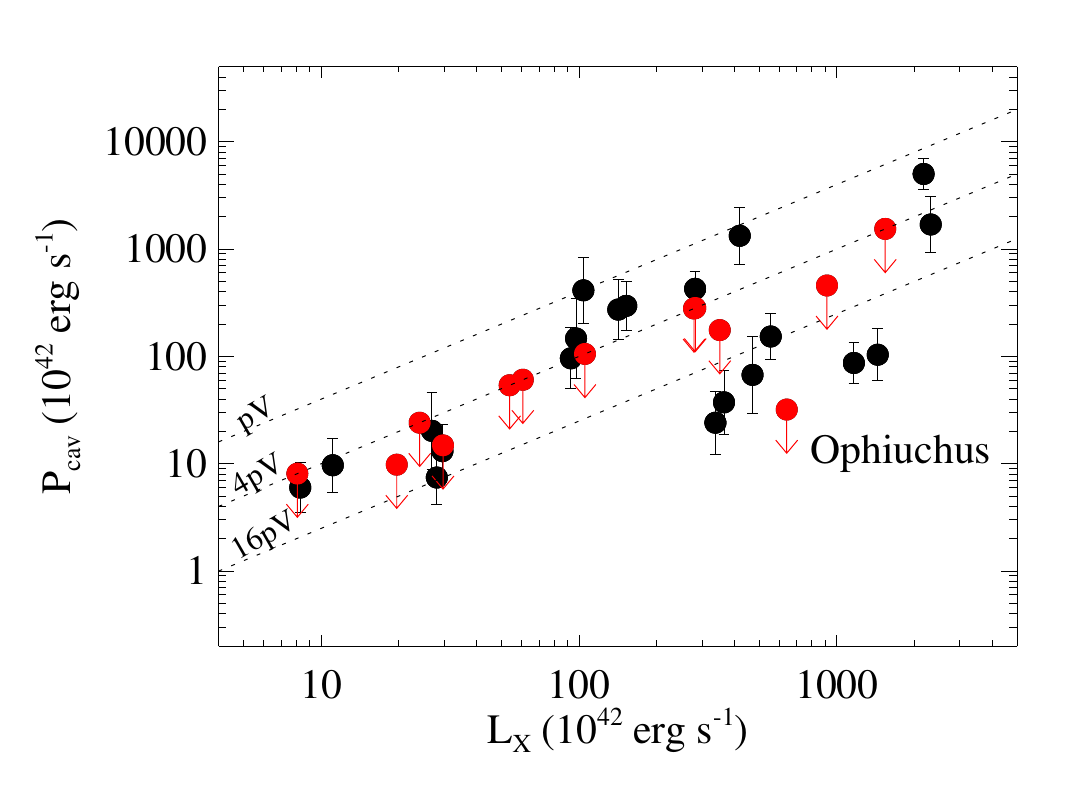}
\end{array}$
  \caption{ \emph{Left}:  Simulated image of A2244 with the bubble size and location calculated assuming adiabatic evolution, with $P_{\rm{cav}} \sim 10L_{\rm{X}}$ and the angle from the line of sight of 90$^{\circ}$. \emph{Right}: Cavity power versus the bolometric X-ray luminosity for 33 cooling flow clusters selected from the B55 sample. The upper limits are the 13 systems that do not show cavities in their atmospheres, the remaining 20 systems have cavities. The cavity power and X-ray luminosity for the systems with cavities are all, except A496, from the literature (see text for details).}
\end{figure}

\bibliographystyle{aipproc}   

\bibliography{../master_references}

\IfFileExists{\jobname.bbl}{}
 {\typeout{}
  \typeout{******************************************}
  \typeout{** Please run "bibtex \jobname" to optain}
  \typeout{** the bibliography and then re-run LaTeX}
  \typeout{** twice to fix the references!}
  \typeout{******************************************}
  \typeout{}
 }

\end{document}